\documentclass[12pt,preprint]{aastex}

\newcommand{\gapprox}   {\lower.4ex\hbox{$\;\buildrel >\over{\scriptstyle\sim}\;$}}
\newcommand{\lapprox}   {\lower.4ex\hbox{$\;\buildrel <\over{\scriptstyle\sim}\;$}}

\newcommand{\begeq}     {\begin{equation}}
\newcommand{\fineq}     {\end{equation}}

\newcommand{\Msun}      {\mbox{$\,M_{\mathord\odot}$}}

\newcommand{\OMC}		{\mbox{${\rm O}-{\rm C}$}}

\newcommand{\Porb}{\mbox{$P_{\rm orb}$}}

\newcommand{\Porbdot}{\mbox{$\dot P_{\rm orb}$}}

\newcommand{\Porbddot}{\mbox{$\ddot P_{\rm orb}$}}

\newcommand{\exo}{\mbox{EXO\,0748$-$676}}
\newcommand{\xonesixfiveeight}{\mbox{X1658$-$298}}

\newcommand{\sgrastar}{\mbox{Sgr\,\,A*}}

\begin{document}

\title{Eclipse Timings of the Transient Low Mass X-ray 
Binary EXO 0748$-$676. IV.
The Rossi X-Ray Timing Explorer Eclipses}

\author{Michael T. Wolff\altaffilmark{1},
Paul S. Ray\altaffilmark{2}, Kent S. Wood\altaffilmark{3}}
\affil{Space Science Division, 
Naval Research Laboratory,
Washington, DC 20375-5352}
\and
\author{Paul L. Hertz\altaffilmark{4}}
\affil{Science Mission Directorate, 
NASA Headquarters, 
Washington, DC 20546-0001}

\altaffiltext{1}{E-mail Address: Michael.Wolff@nrl.navy.mil}
\altaffiltext{2}{E-mail Address: Paul.Ray@nrl.navy.mil}
\altaffiltext{3}{E-mail Address: Kent.Wood@nrl.navy.mil}
\altaffiltext{4}{E-mail Address: Paul.Hertz@nasa.gov}

\begin{abstract}

We report our complete database of X-ray eclipse timings 
of the low mass X-ray binary \exo\ observed by the 
{\it Rossi X-Ray Timing Explorer} ({\it RXTE}) satellite.
As of this writing we have accumulated 443 full X-ray eclipses, 
392 of which have been observed with the Proportional Counter
Array on {\it RXTE}.
These include both observations where an eclipse was specifically 
targeted and those eclipses found in the RXTE data archive.
Eclipse cycle count has been maintained since the discovery of 
the \exo\ system in February 1985.
We describe our observing and analysis techniques for each eclipse
and describe improvements we have made since 
the last compilation by \citeauthor{whw+02}
The  principal result of this paper is the database containing 
the timing results from a seven-parameter fit
to the X-ray light curve for each observed eclipse along with the 
associated errors in the fitted parameters.
Based on the standard \OMC\ analysis, \exo\ has undergone four 
distinct orbital period epochs since its discovery.
In addition, \exo\ shows small-scale events in the \OMC\ curve that 
are likely due to short-lived changes in the secondary star.

\end{abstract}


\section{Introduction}
\label{section:introduction}

The orbital period is one of the most fundamental 
parameters characterizing binary star systems.
Understanding the evolution of binary systems requires 
understanding how the orbital period might change on both 
short and long time scales, both from a theoretical and 
observational standpoint.
As a binary system evolves mass and 
angular momentum are exchanged between the binary components,
angular momentum is carried out of the system via mass loss, and
each component can evolve via either nuclear compositional and
stellar structural changes.
Eclipse transitions provide good fiducial timing markers allowing the 
precise timing of the orbit which, in turn, make possible the 
long-term characterization of period changes and the 
magnitude of the effects these physical processes have on
the system parameters.

To date, five low mass X-ray binaries (LMXBs) showing full X-ray 
eclipses are known (\exo, \xonesixfiveeight, XTE$\,$J1710$-$281, 
AX$\,$J1745.6$-$2901, and GRS$\,$J1747$-$312).
Of these sources, only \exo\ has been persistently X-ray active
over the past two decades and thus available for continuous 
monitoring of its orbital period behavior \citep{pwgg86,whw+02}.
\exo\ is a transient X-ray system with a 3.82 hour orbit period 
first observed by the European Space Agency's X-ray Observatory 
({\it EXOSAT}) satellite when it became X-ray active 
in February of 1985 \citep{pwgg86}.
It has remained X-ray active at a level of at least a few mcrab 
since its discovery. 
The eclipsing LMXB \xonesixfiveeight\ was discovered by the SAS-3 
satellite in October 1976 \citep{lhdl76} and was eventually 
determined to be a X-ray bursting, eclipsing LMXB with
an orbital period of 7.12 hours \citep{cw89}.
\xonesixfiveeight\ became undetectable in X-rays in 1979 but then 
returned to X-ray activity in April 1999, repeatedly being 
observed by BeppoSAX \citep{ops+01} and the
{\it Rossi X-Ray Timing Explorer} ({\it RXTE}) \citep{wsb00}
until it became X-ray inactive again in October 2000.
The observations reported by \citet{wsb00} appeared to indicate 
that the orbital period was changing with a time scale 
of $\tau_{orb} = | \Porb / \Porbdot | \sim 10^7$ years
even though little mass transfer had occurred between epochs
of X-ray activity.
\xonesixfiveeight\ has remained X-ray inactive since
October 2000 making long term monitoring of its X-ray 
orbital period impossible.

GRS$\,$J1747$-$312 in globular cluster Terzan 6, and 
XTEJ1710$-$281 are both regularly recurring transient sources 
showing full X-ray eclipses and have orbital periods 12.4 hours 
and 3.28 hours, respectively.
Their ephemerides can be established via X-ray 
observations \citep{zhm+03}. 
However, in the case of GRS$\,$J1747$-$312, timing observations based on
the mid-eclipse timings are difficult because of the relatively long
eclipse totality ($\sim43$ minutes).
Eclipse observations must wait through a long duration of totality when 
virtually no X-ray activity from the source can be seen in order to 
accurately time both the ingress and egress transitions of any  
single eclipse. 
Finally, the transient faint source AX$\,$J1745.6$-$2901 is very
close to \sgrastar\ and thus very difficult to observe 
because of source confusion in that crowded field. 
Nevertheless, it has been detected by both the Advanced Satellite 
for Cosmology and Astrophysics ({\it ASCA}) \citep{mks+96,skm+02} 
and by the X-ray Multi-Mirror Mission ({\it XMM-Newton}) with 
the 23-minute eclipse variations being visible and the 
orbital period measured at 8.36 hours \citep{pgg+07}.

For all of these reasons, \exo\ is a uniquely good system for 
long-term study of LMXB orbital evolution.
Thus, early in the {\it RXTE} mission, we began a program of 
monitoring the X-ray eclipses in \exo\ in an 
effort to detect orbital period variations. 
The ultimate goal at that time was to measure the orbital period
derivative \Porbdot\ and compare this with estimates of
magnitude of orbital period changes in 
theoretical calculations of LMXB evolution.
\exo\ was observed intensely by the EXOSAT satellite \citep{pwgg86}
resulting in the first estimate of its orbital period and also
revealing that, because the source showed X-ray bursts, 
the compact object in the binary was a neutron star.
It should be noted, however, that even the 23-year time baseline
reported on here is short compared to the time steps in most 
evolutionary calculations of binary systems which are in the 
range $\Delta t > 10^{4-5}$ years.
Theoretical calculations of LMXB evolution must cover
a time span of 1-10 Gyrs and time-steps less than $10^4$ years 
would make such computations prohibitively expensive.
Thus, we still have yet to access a sufficiently long baseline
to allow robust comparison with most theoretical calculations of 
LMXB evolution. 
Conversely, most theoretical calculations do not resolve binary
system effects, such as magnetic field generation and cycling in the 
secondary star, that can potentially influence the system \OMC\ behavior 
on our 23-year observational time scale.

With these monitoring observations we expected to find smooth 
variations in the \OMC\ residuals (observed minus calculated
residuals are the delay or advance in the  
mid-eclipse timings compared to that expected for a 
specified ephemeris) that could be compared to 
the estimates of LMXB evolutionary time scales predicted by 
theory \citep[e.g.,][]{rjw82}.
What we found was considerably more complicated.
The first paper in this series, \citet{hwc95}, concluded that
the variable eclipse durations and profiles observed in \exo\ 
implied there is an additional source of uncertainty in timing 
measurements and that this uncertainty is intrinsic to the binary 
system, correlated from observation to observation, and had 
variance that increased as a function of binary cycles 
between observations.
\citeauthor{hwc95} also suggested that spurious trends in the \OMC\
residuals could be misinterpreted as changes in the orbital period.
\citet{hwc97} went further and identified a component of 
``intrinsic scatter'' in the \OMC\ residuals of $\sim 0.15$ s per 
orbit cycle.
Finally, \citet{whw+02} found that neither a constant orbital 
period derivative or any reasonable polynomial expansion 
in \Porb, \Porbdot, etc., provided an acceptable fit to the 
entire historical set of \OMC\ residuals.
Rather, the system appeared to go through ``epochs'' in which 
the orbital period would change on the order of milliseconds 
from epoch to epoch.
A model for the observed \OMC\ variations that included 
observational measurement error, cumulative orbital period 
jitter intrinsic to the binary system, with some underlying 
period evolution was found to be consistent with the observed 
mid-eclipse timing data and implied a rapid timescale for 
orbit evolution, $\tau_{orb} \sim 2 \times 10^7$ years. 

\section{Full Eclipse Observations}
\label{section:observations}

\subsection{{\it RXTE} Observations}
\label{section:rxteobs}

We utilize all observations from the {\it RXTE} satellite \citep{jsg+96,jmr+06} 
of \exo\ from which we can isolate a full X-ray eclipse. 
We number the eclipses using the numbering scheme of \citet{pwgg86} but 
with a revised ephemeris: 
$T_0 ({\rm TDB;MJD}) = 46111.0751315$ and $\Porb = 0.1593377866$ days. 
We reduced each observation data set with the HEASoft software 
version 6.5 released on June 26, 2008. 
We utilize the combined background model released on
August 6, 2006 available at the {\it RXTE} web
site\footnote{PCA X-ray background model information is
found at the web site http://rxte.gsfc.nasa.gov/docs/xte/pca\_news.html.}.
These software releases are corrected for the problems 
with the background estimation algorithms and 
the Proportional Counter Array (PCA) effective area calibration 
that makes the Crab flux come out at its 
standard value \citep[see][]{jmr+06}.

The observations of \exo\ we analyze generally fall into 
two categories. 
First, there are our systematic monitoring observations or
``campaigns''. 
These are tightly clustered observations over about one day 
of a series of eclipses, perhaps consecutive, 
made around the approximate predicted eclipse times, giving us a 
look at the pre-eclipse flux level for several hundred seconds, 
the full eclipse profile, and finally the post-eclipse flux 
level, again for several hundred seconds.
The duration of X-ray eclipse in \exo\ is approximately 495 seconds
($\sim$8.3 minutes) and so a complete observation consists
of as little as $\sim 1.5$ ks of PCA time.
This makes \exo\ a good target for eclipse 
timing compared to, say, GRS1747-312, with its 43 minute 
totality duration.
Virtually all of our monitoring eclipse observations,
except those early in 1996, where made with the GoodXenon 
Experiment Data System (EDS) mode.
Early 1996 observations, however, done before we had much 
experience with the PCA response to X-ray sources, were made with 
the E\_62us\_64M\_0\_1s, E\_8us\_32B\_0\_1s, 
E\_125us\_64M\_0\_1s, E\_250us\_128M\_0\_1s EDS modes.
The second general type of observation are those observations 
of \exo\ that were done for other scientific reasons such as to 
probe quasi-periodic oscillations or searching for X-ray bursts 
and are found in the High Energy Astrophysics Science 
Archive Research Center (HEASARC) 
public archive at Goddard Space Flight Center.
Generally, these observations are done in an event mode
such as E\_125us\_64M\_0\_1s.
 
Early in the {\it RXTE} mission our campaign observations,
indeed most observations of any source, utilized the
full complement of 5 Proportional Counter Units (PCUs).
As the problems developed over time with various PCUs the
number of PCUs in any one observation was reduced 
so that by the end of the {\it RXTE} mission it is not unheard of
for only one or two PCUs to be turned on for a particular observation.
In the case of GoodXenon data both layer and PCU information
is preserved in the raw data files we start with and thus 
we can select events based on PCU and anode layer within each PCU. 
On the other hand, generally for the event modes, 
layer information (and sometimes PCU information) was not 
available in the raw data files. 

During most observations the background count rates in one 
PCU for layer 1 in the energy range 2-20 keV 
are around 7-10 counts s$^{-1}$ PCU$^{-1}$.
If the observation mode is GoodXenon we select events occurring 
in layer 1 from those PCUs on during the entire eclipse observation 
thus keeping the background count rate low.
On the other hand, if we must use events from all layers
then typical background count rates in the same 
energy range will be significantly larger, increasing the 
Poisson errors and  
making the eclipse light curves more noisy.
If a PCU turned on or off during an eclipse this would lead
to variations in the background level and source count rates that
invalidated our model for the simple ingress/egress fitted shape of 
the eclipse.
Furthermore, when a PCU is turned on or off during an observation 
this can lead to count rate transients that interfere with 
our eclipse fitting. 
We screened our observations for such effects and when they
were found we first tried to recover a fitable eclipse profile by
excluding events from the
PCU that was turned on/off during the observation.  
If for any reason it was not possible to exclude only the 
offending PCU then we excluded the entire eclipse.
Finally, we note that we make no effort to exclude 
either of the PCUs that have no Propane layer but instead simply
accept the events from those PCUs and utilize
the existing background estimates from {\it pcabackest}.

We proceeded as follows for our analysis of each eclipse: 
We extracted light curves as described above for each eclipse,
screening the observations for electron or other transient
instrument events that can sometimes degrade PCA performance, and when
such problems did occur and they affected the entire light curve 
we would exclude that eclipse from our analysis. 
In rare instances short data dropouts
can occur during an observation that show up as either gaps
in the final source light curve or as 2-second segments
for which, regardless of the true source plus background count rate, 
the count rate goes to zero.
In general, we mask such regions out of the analysis by selecting
events outside the relevant time intervals.
However, we exclude the entire eclipse from analysis if such a 
corrupted region in the light curve affects a transition feature 
(ingress or egress) in the fitted light curves.
In those cases where there was an 
X-ray burst either during totality or in the parts of the 
uneclipsed light curves we fit to, we masked out the part of 
the light curve containing visible signs of the burst in order not 
to bias the estimate of the normal X-ray count rate.
The only exception to this was when a burst occurred sufficiently
close to either the ingress or egress to distort 
the shape of the transition.  
In this case we excluded the eclipse from further analysis.
The background was estimated with the FTOOL {\it pcabackest}
utilizing the faint-source model, 
the resulting source+background and background light curves 
were barycentered with the FTOOL {\it faxbary}, and then the 
background was subtracted from the source+background light 
curve to produce a source-only barycentered lightcurve 
showing the full eclipse. 
The barycentering FTOOL {\it faxbary} has an absolute timing 
accuracy of better than 100$\mu$s which is sufficient for 
our purposes \citep[see][]{rjm+98}.
This resulting source-only light curve, usually binned at 0.5 s, 
was then fitted with our seven-parameter ramp-and-step model for 
the basic eclipse variations. 

\subsection{Eclipse Fitting and Error Estimates}
\label{section:eclipsefitting}

Our model for a full X-ray eclipse consists of seven parameters: 
the pre-ingress ($F_{in}$), totality ($F_{ec}$), and post-egress ($F_{eg}$)
count rates (uncorrected for number of PCUs), 
the durations of the ingress (${\Delta t}_{in}$) and egress 
(${\Delta t}_{eg}$), the duration of totality (${\Delta t}_{ec}$) 
and finally the barycentered mid-eclipse time ($t_{mid}$).
An alternative model for the eclipse times is the 
times of the four ``contacts'': $t_1$, $t_2$, $t_3$, and $t_4$. 
The relationship between the two timing systems is simple: 
\begin{eqnarray}
t_1  & = & t_{mid} - \frac{1}{2} {\Delta t}_{ec} - {\Delta t}_{in} , \\
t_2  & = & t_{mid} - \frac{1}{2} {\Delta t}_{ec} , \\
t_3  & = & t_{mid} + \frac{1}{2} {\Delta t}_{ec} , \\
t_4  & = & t_{mid} + \frac{1}{2} {\Delta t}_{ec} + {\Delta t}_{eg} .
\end{eqnarray}

\begin{figure}
\centerline{\includegraphics[height=5.9in,angle=270.0]{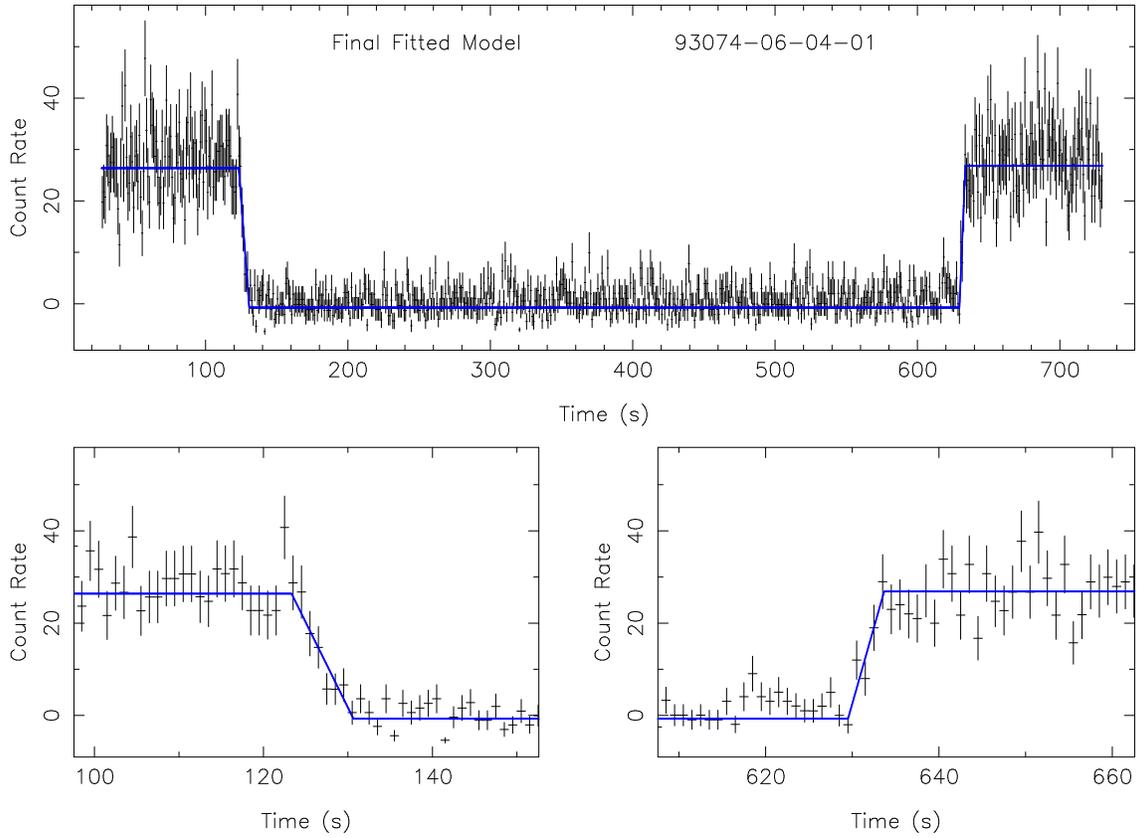}}
\caption{A seven-parameter full X-ray eclipse model (blue line) fitted 
to the X-ray eclipse light curve from the indicated observation.
This is a typical fit to one of our eclipse observations. 
The top panel shows the full light curve and model fit while the
two bottom panels show the details of the fits to the ingress (left) and
the egress (right) transitions.
\label{fig:eclipsefit}}
\end{figure}

A fitted eclipse from an observation utilizing only
one PCU is shown in Figure~\ref{fig:eclipsefit}. 
The eclipse light curve fit results in the parameters with error 
estimates that are listed for each eclipse in Table~\ref{tbl:fitable}, and 
constitutes the principal result of this paper. 
All of the errors quoted in Table~\ref{tbl:fitable} come from this
fitting process. 
We also show in the table the fit reduced $\chi^2$ value ($\chi_{\nu}^2$) 
and the number of PCUs utilized during each observation.
We compile a second table of those eclipses or partial eclipses 
observed by {\it RXTE} that we can not fit with our seven-parameter 
model in Table~\ref{tbl:unfitable}, each for the reason indicated.

The fitted mid-eclipse time in Table~\ref{tbl:fitable} is 
simply $t_{mid} = (t_2 + t_3)/2$, the half-way point between 
the end of ingress and the beginning of egress.
When the ingress and egress profiles are relatively smooth 
and the count rate changes monotonically, even in those 
cases where the ingress or egress durations are relatively long, 
we find the mid-eclipse time can be determined with a 
typical accuracy of $\sim1/2$ s. 
If, on the other hand, the ingress or egress changes are not 
monotonic, such as is the case with the momentarily reversing 
ingress profiles discussed in \citet{wwr07}, then determining 
the mid-eclipse times becomes more difficult because the applicability 
of our seven-parameter eclipse model becomes problematic. 
In such a case a more sophisticated model for the eclipse variations
must be employed and the errors must be determined in
some new fashion that will depend on the model chosen.
This is a point to which we return below.

\begin{figure}
\centerline{\includegraphics[height=5.9in,angle=0.0]{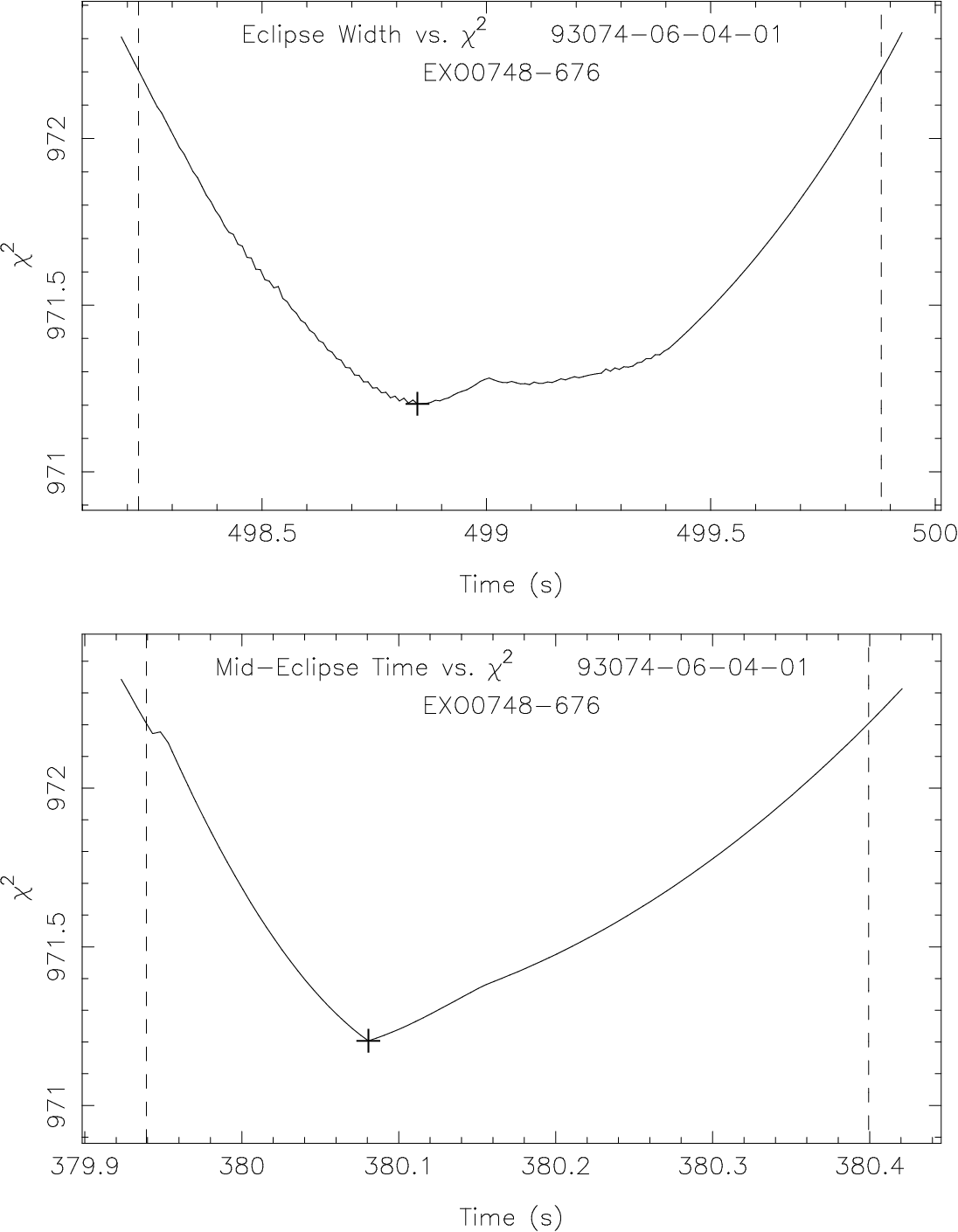}}
\caption{$\chi^2$ variation as a function of eclipse width (top panel) 
and mid-eclipse time (bottom panel) for the fit to the light curve 
from the observation shown in Figure~\ref{fig:eclipsefit}. 
The bottom panel shows the $\chi^2$ variation  
with displacement away from the best fit mid-eclipse time and the 
top panel shows the $\chi^2$ variation with displacement away from 
the best fit eclipse duration.
The best fit value of $\chi^2$ is denoted by the 
cross on both plots.
The dashed lines show where $\chi^2$ crosses a value equal to 
the minimum value incremented by 1 for both positive and 
negative parameter variation.
\label{fig:fitchi2}}
\end{figure}

During the fit process, in order to analyze the \OMC\ behavior 
of the system carefully [e.g., as was done in \citet{whw+02}], we 
must make a robust estimate of the errors in the fitted mid-eclipse times.
A careful analysis of the errors in all seven parameters in as 
analytically rigorous as possible would be very computationally complex 
and too costly in outlay of effort relative to the quality of the 
resulting error estimates.
Thus, we employ a less formally rigorous but more computationally 
tractable procedure to estimate the errors in
the fitted parameters listed in Table~\ref{tbl:fitable}.
First, we utilize a very careful algorithm to find the best 
fit eclipse model and global minimum $\chi^2$ value in 
the seven-parameter fit space for our eclipse model.
Once we have the best fit model parameters and ${\chi^2}_{min}$ we 
then carefully search in the mid-eclipse time parameter $t_{mid}$
around this ${\chi^2}_{min}$ until we have those 
values of the mid-eclipse time both above and 
below $t_{mid}$ for which $\chi^2$ increases by 1 over ${\chi^2}_{min}$.
This gives us two time intervals, one above $t_{mid}$ and one
below $t_{mid}$, representing the error time interval for $t_{mid}$
and it is the average of these two intervals that we give
as the error in $t_{mid}$ in Table~\ref{tbl:fitable} column 4.
We employ a similar procedure to estimate the errors in 
the eclipse widths (${\Delta t}_{ec}$) in Table~\ref{tbl:fitable}.
The results of this process are illustrated in Figure~\ref{fig:fitchi2}
for an example eclipse where the pre- and post-eclipse count rate is 
relatively low, in part because only one PCU was employed 
during the observation.
However, in order to save computational complexity, we 
crudely estimate the errors in the other fit parameters based on
the square root of the covariance matrix diagonal elements:
$\sigma_{a_i} \, \sim \, \sqrt{cov(a_i,a_i)}$ 
where $a_i$ is the parameter of interest and $cov(a_i,a_j)$ is 
represents the element of the fit covariance matrix for 
parameters $a_i$ and $a_j$.
Thus, we are implicitly assuming that the variation in $\chi^2$ 
for the other five parameters in the fit are parabolic in the 
region around ${\chi^2}_{min}$ \citep[e.g.,][]{br03}.

\section{Observed Ephemeris Behavior}
\label{section:ephemeris}

\begin{figure}
\centerline{\includegraphics[height=5.9in,angle=270.0]{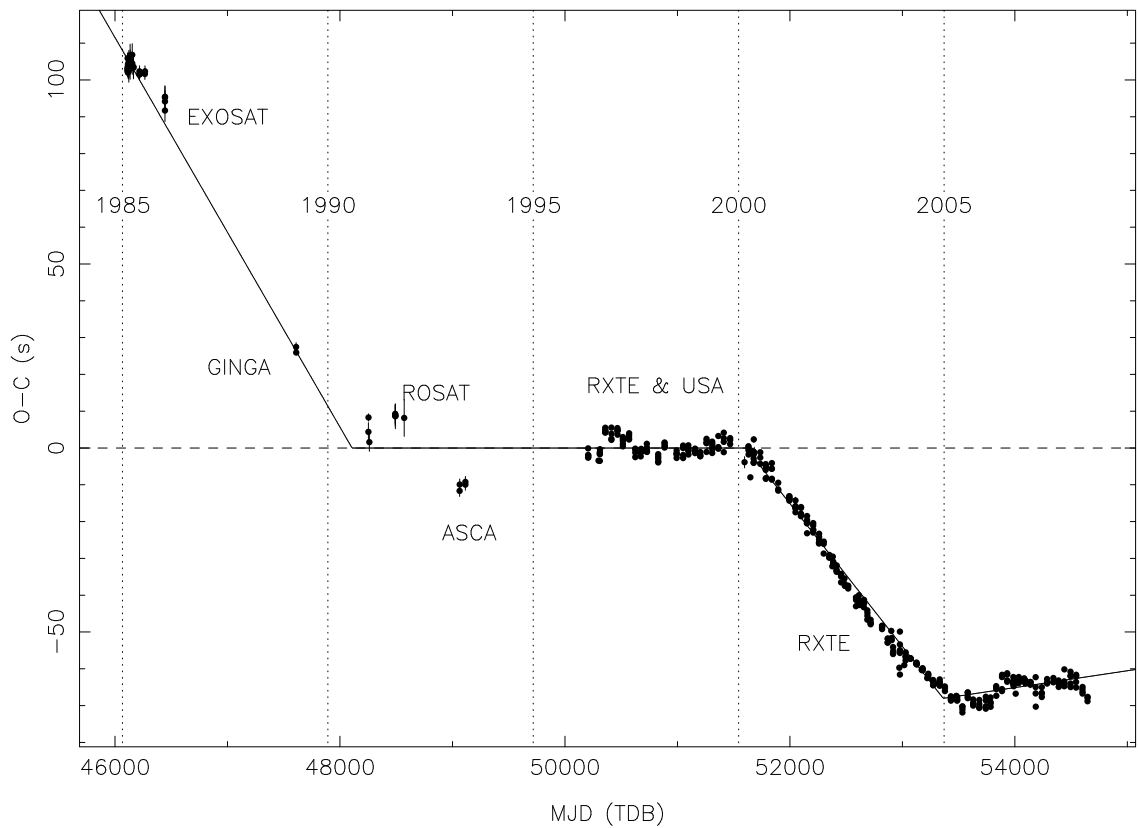}}
\caption{The mid-eclipse timing residuals for observed eclipses 
of \exo\ during 1985$-$2008 both from the present study and 
those available in the literature.  
The residual of the observed mid-eclipse time is plotted as a
function of barycenter corrected observation date. 
The curved solid line is the four-constant period  
solution to all the data described in the text. 
No simple linear or quadratic ephemeris fits all the data
points.
\label{fig:omcalldata}}
\end{figure}

Figure~\ref{fig:omcalldata} shows the \OMC\ diagram for the entire 
eclipse database from {\it EXOSAT}, the GINGA X-ray satellite, {\it ASCA}, 
the Roentgen Satellite ({\it ROSAT}), 
and the Unconventional Stellar Aspect experiment ({\it USA}).
The sign convention for the \OMC\ is such that \OMC\ will
be positive for an observed mid-eclipse time coming after
the calculated (ephemeris predicted) time.
No simple polynomial expansion in \Porb, \Porbdot, \Porbddot, etc., 
can fit the \OMC\ variations apparent in the figure.
However, four epochs of general \OMC\ behavior are evident from 
Figure~\ref{fig:omcalldata}.
A piecewise linear function of mid-eclipse time 
representing the orbital period in the four distinct epochs can 
approximately represent the general trends in the \OMC\ curve.
In such a model we constrain the orbital phase to be constant across
the instantaneous period change but allow the cycle of the change 
to be a free parameter.
Such a function can be written in the form: 
\begin{equation}
T_n  =
\cases{
T_0 + n \Porb_{,0} &{if $n \leq n_{b,0}$},\cr
T_0 + n_{b,0} \Porb_{,0} + ( n - n_{b,0} ) \Porb_{,1} &{if $n_{b,0} < n \leq n_{b,1}$},\cr
T_0 + n_{b,0} \Porb_{,0} + ( n_{b,1} - n_{b,0} ) \Porb_{,1} + ( n - n_{b,1} ) \Porb_{,2} &{if $n_{b,1} < n \leq n_{b,2}$},\cr
T_0 + n_{b,0} \Porb_{,0} + ( n_{b,1} - n_{b,0} ) \Porb_{,1} + ( n_{b,2} - n_{b,1} ) \Porb_{,2} + ( n - n_{b,2} ) \Porb_{,3} &{if $n_{b,2} < n$}.\cr
} 
\end{equation}
where the eclipse cycle number is $n$, the $n_{b,i}$ are the
cycle numbers where transitions from one period to another period are
made, and the orbital periods in each epoch are $P_{orb,i}$.
Table~\ref{tbl:4epochephem} gives the results of such a piecewise fit
to all the mid-eclipse times for \exo. 
This solution for the orbital period is only an approximate 
representation of the \OMC\ behavior on several year time 
scales and the formal $\chi^2$ value for the fit is very poor.  
The individual \OMC\ residuals are often clumped either above or 
below the solution line in the figure indicating that variation 
in the orbital period is occurring that our four-epoch solution 
does not capture.
Thus, no formal errors are quoted in Table~\ref{tbl:4epochephem} for 
the parameters of the fit to equation 5.
The systematic wandering of the mid-eclipse timings around
the mean period noted by \citet{hwc97} and \citet{whw+02} has 
continued for all the eclipses observed by {\it RXTE}.

\begin{figure}
\centerline{\includegraphics[height=5.9in,angle=270.0]{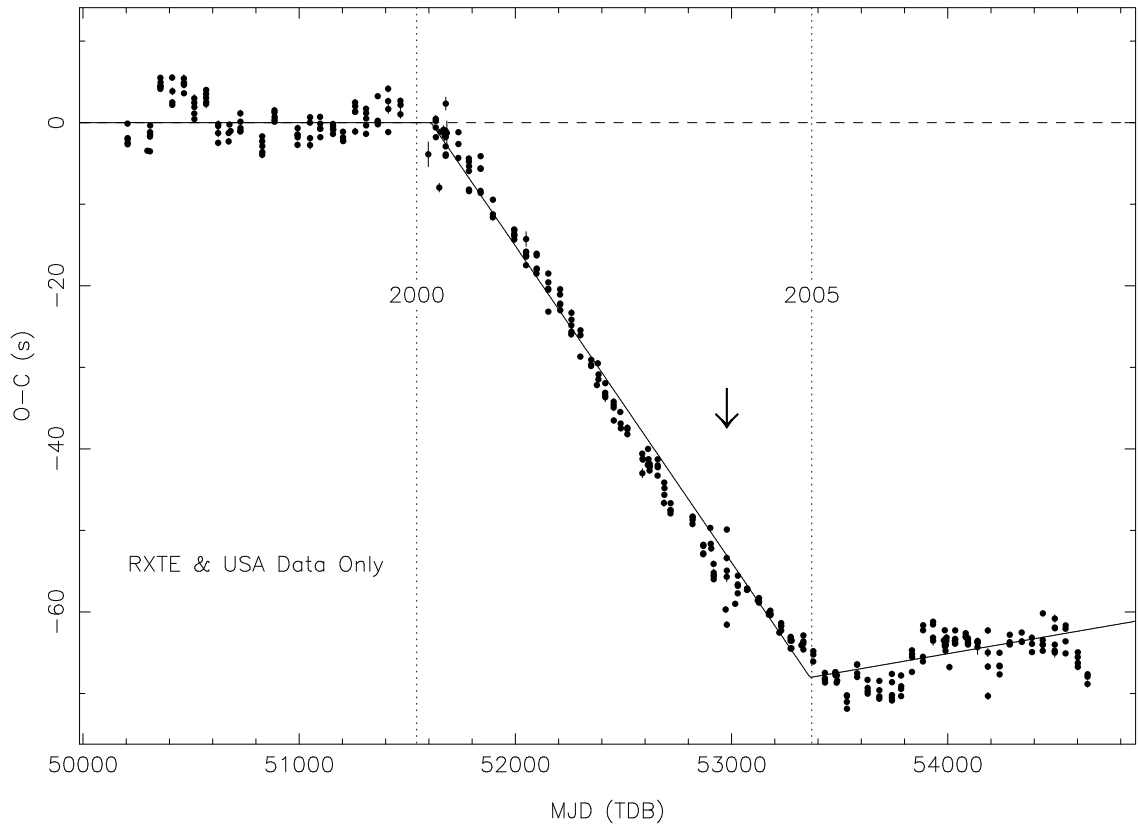}}
\caption{The mid-eclipse timing residuals for observed eclipses 
of \exo\ during 1996$-$2008 (the {\it RXTE} era).  
The solid line represents the four-constant period  
solution to all the data.
The large variations in \OMC\ values away from the solid line
solution at MJD 52978 are the ``magnetic loop'' eclipses described
in \citet{wwr07}.
\label{fig:omcrxtedata}}
\end{figure}

Based on our four-epoch solution to the \OMC\ variations 
shown in Figure~\ref{fig:omcalldata} it
is apparent that at least four distinct orbital period epochs 
have occurred in the \exo\ system since its discovery in 1985.
The three epochs that have occurred during the {\it RXTE} mission 
alone are shown in Figure~\ref{fig:omcrxtedata} and this figure 
also includes seven eclipses observed by the {\it USA} 
experiment that operated from May 1999 to November 2000.
One of these orbital period epochs is completely 
sampled by the {\it RXTE} observations, namely, the 4.8 years 
from MJD 51611 to MJD 53363.

\citet{wwr07} reported a consecutive series of eclipses where the
occulting edge of the secondary star appeared to be modified from a 
simple hydrostatic atmosphere by a magnetically confined 
loop suspended above the secondary star's surface. 
In their model the loop absorbed X-rays from the neutron star 
accretion disk just before ingress casting a X-ray shadow as 
seen at the earth and modifying the normal ingress transition 
profile.
The five eclipses involved in that event are shown by the 
arrow at MJD 52978 in Figure~\ref{fig:omcrxtedata}.
It is precisely for such eclipses, where the ingress or egress 
profiles consist of other than simple smooth, 
monotonically changing count rates, that our seven-parameter 
model for the eclipse profile breaks down and the fitted mid-eclipse 
time errors are no longer reliable. 
However, even in such cases our simple fits can yield the useful 
information of how such anomalous eclipses are bunched together 
over time, i.e., how much the system ``remembers'' the anomalous 
transitions from eclipse to eclipse \citep{koen96,hwc97,whw+02}.

\begin{figure}
\centerline{\includegraphics[height=5.9in,angle=270.0]{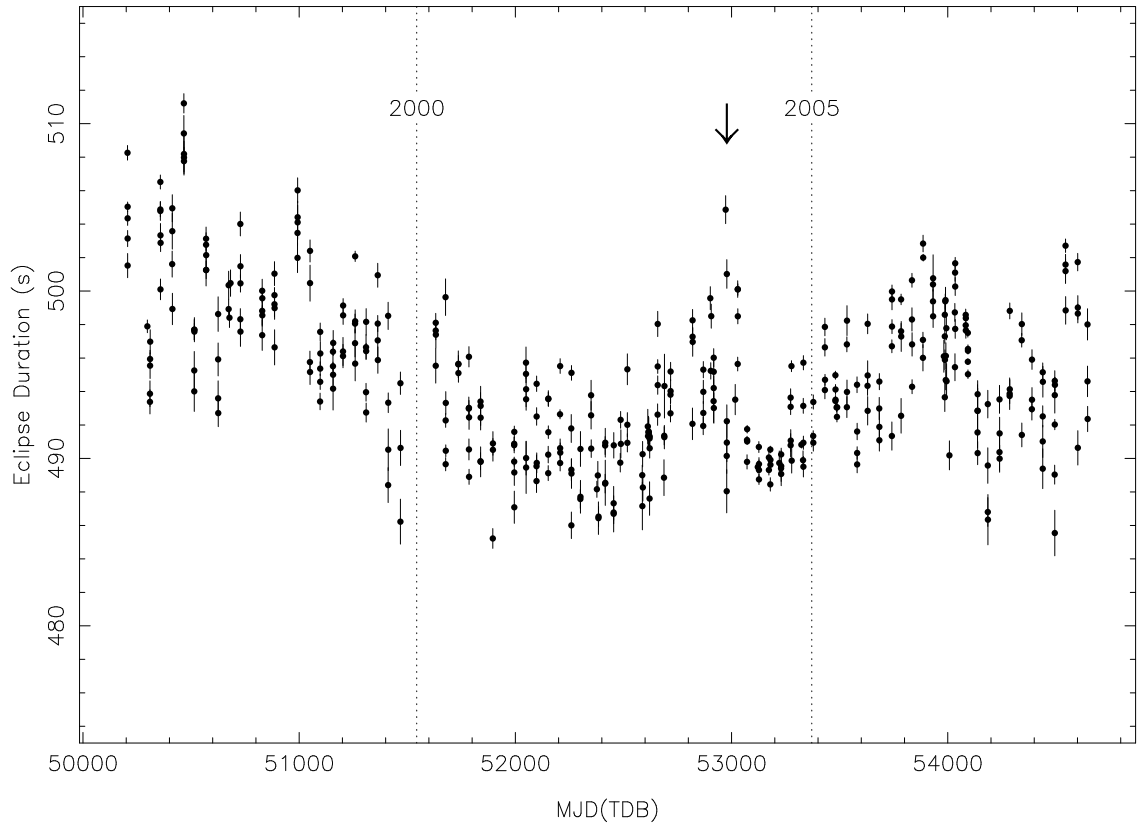}}
\caption{
The duration of totality (${\Delta t}_{ec}$) for {\it RXTE} fitted eclipses 
of \exo\ during 1996$-$2008 from Table~\ref{tbl:fitable}.
Early in the {\it RXTE} mission eclipse duration appears to 
shorten as the secondary star undergoes a general contraction.
However, near MJD 52000 this trend is reversed as the 
secondary expands somewhat. 
The arrow shows the time of the disruptions of the
\OMC\ residuals shown in Figure~\ref{fig:omcrxtedata}.
Note also the stable eclipse durations ($\sim490$ s) 
around MJD 53200 that have small associated error estimates.
\label{fig:rxteduration}}
\end{figure}

In their discussion of the magnetic eclipses, \citet{wwr07}
stated that roughly 93 days after the first set of anomalous 
eclipses the eclipse profiles became particularly 
sharp and stable. 
The evidence supporting this claim can clearly be seen in 
Figure~\ref{fig:rxteduration} where
we plot the duration of eclipse (${\Delta t}_{ec}$) as a function 
of time for the Table~\ref{tbl:fitable} eclipses.
Analysis of the anomalous eclipses studied by \citet{wwr07}
reveals large variations in the \OMC\ residuals,
due primarily to the long durations in the ingress times.
However, eclipses in the range MJD $= 53071 - 53229$ 
have shorter durations ($\sim$490 seconds) and 
significantly smaller mid-eclipse time and eclipse duration 
uncertainties when compared to the eclipses on either side 
of this MJD range.
During the eclipses analyzed by \citeauthor{wwr07} 
the secondary was in a more extended state with the radius
of its X-ray occulting edge at a larger value than the 
more stable eclipses coming $\sim$93 days later. 

Examination of Figure~\ref{fig:omcalldata} strongly suggests that
the \exo\ system is currently undergoing cyclic behavior. 
The character of the \OMC\ residuals strongly resemble 
similar behavior of Algol binary 
systems \citep[e.g.,][]{sode80,simo97}.
Numerous studies have been done of the \OMC\ variations in
Algol systems and the currently favored model is the 
\OMC\ variations are brought about by magnetic cycling in
the one of the binary components, in some ways similar to 
the 22-year solar magnetic cycle \citep{hall90}.
The orbital period changes we are seeing in \exo, 
$ | {\Delta \Porb} / \Porb | \sim 5 \times 10^{-7}$, are 
similar in magnitude to the observations of orbital period changes 
in Algol and RX CVn binaries \citep[e.g.,][]{hall90}.
If the observation of magnetic loop structures \citep{wwr07} 
in the chromosphere of the secondary star in the \exo\ 
system is correct, then a magnetic cycling model for 
the \OMC\ variations in \exo\ gains strong circumstantial support.
Before the {\it RXTE} era X-ray eclipse observations did not 
capture all the variations in the \OMC\ residuals.
Thus, no statement can be made about magnetic cycle time
scales before the onset of {\it RXTE} observations.
Once {\it RXTE} observations commenced in 1996, however,
the richness of the \OMC\ variations became apparent.
The timescale for one cycle can be estimated as
approximately twice $T_{b,2} - T_{b,1} \sim 4.8$ years from 
Table~\ref{tbl:fitable}, or $\sim$9.6 years for a complete cycle 
of the secondary star's magnetic field {\it if} the
cyclic field variations are similar to the polarity reversals
experienced by the sun during one 22-year magnetic cycle.
Exact predictions regarding when the system orbital period 
will change again are difficult, however, because we do not 
know exactly when the period shifted between $\Porb_2$ and $\Porb_3$.
The \OMC\ residuals near $n_{b,2}$ suggest that a gradual
change in orbital period may have occurred.

In order to account for cyclic changes in the \exo\ 
orbital period, we must consider changes that occur 
on a time scale significantly less than the time scale 
for either mass exchange, spin-orbit coupling,
or orbital circularization between the neutron star and secondary. 
We summarize here the model of \citet{lrr98} and \citet{lr99} where 
a detailed development of the magnetic cycling and oribtal 
period modulation theory is given. 
We assume that the \exo\ binary consists of a magnetically
active secondary star moving in the gravitational field of the
neutron star, that the orbits are circular, and that the 
equitorial plane of the secondary lies roughly in the plane 
of the orbit. 
The secondary star in the \exo\ system is near 0.4 \Msun\ 
implying that there is likely a convective envelope in
the star and a resulting stellar magnetic field.
The orbital angular momentum of the system will depend on the 
orbital radius, the orbital velocity and the mass of each stellar 
component. 
However, the cyclic changes of the magnetic field of the secondary 
induces cyclic changes in the secondary's structure as magnetic 
pressure support in the convective layer varies.
Thus, the gravitational quadrupole moment of the secondary star 
can be time-dependent and induce variable orbital motion. 
Since the mass of each component and orbital angular momentum are 
constant on the decadal time scales of interest here the orbital 
radius and the orbital velocity must undergo compensating changes as
the gravitational field changes. 
But if the effective orbital radius changes, then, via Kepler's law, 
the orbital period must also change. 
When the secondary's gravitational quadrupole moment increases, the 
the gravitational pull from the secondary must also increase. 
This, in turn, will cause the two binary components to move toward each 
other causing the orbital velocity to increase and the orbital period 
to decrease.
Conversely, as the quadrupole moment decreases the opposite effect
occurs, again at constant orbital angular momentum, and the 
orbital period increases. 
All of this occurs on a time scale set by variations in the secondary's 
magnetic field and not by the circularization or tidal coupling time 
scales that are expected to be significantly longer for
such systems, and also not by the system mass exchange time scale.
In a future publication we will report on a detailed comparison of 
the magnetic activity model with the \exo\ system behavior. 

\section{Conclusions}
\label{section:conclusions}

We have analyzed 392 full X-ray eclipses from the \exo\ system 
observed by the {\it RXTE} satellite.
We have carefully fitted a seven-parameter model to each 
eclipse light curve profile in order to determine the 
mid-eclipse timings and the errors on those timings.
The observed \OMC\ behavior of the \exo\ system we have found is 
much different than was expected when this project was started. 
The evolution of the binary system after more than 12
years of {\it RXTE} observations and eclipse timings is one of 
multiple orbital period epochs where the orbital period takes 
on distinct values that change rapidly.
We have identified magnetic field cycling of the secondary star
as the most likely cause of these \OMC\ residual variations
and given a brief description of how variations in the
secondary's magnetic field might bring about changes in the
system orbital period.
Also, the \OMC\ diagram shows significant intrinsic jitter 
of various magnitudes during each separate epoch. 
This jitter most likely reflects
subtle changes in the occulting edge of the secondary star
caused by magnetic eruptions in the secondary star's chromosphere.

\begin{acknowledgements}

We thank Jean Swank, Keith Jahoda, Alan Smale, Jacob Hartman,  
Deepto Chakrabarty, Mark Strickman, Neil Johnson, Peter Becker, 
Steve Howell, Craig Markwardt, and Philipp Podsiadlowski for 
important discussions.
We thank Dr. Jeroen Homan for allowing us to look at proprietary
data in advance of publication.
We thank Evan Smith for invaluable help in 
scheduling the eclipse observations with {\it RXTE}. 
We thank an anonymous referee for a number of suggestions 
that helped to improve this paper.
This work was supported by NASA Astrophysical 
Data Analysis Program, the NASA {\it RXTE} Guest Observer Program, 
and by the Office of Naval Research.

\end{acknowledgements}


\clearpage

%
%
\begin{deluxetable}{cccccccccccc}
\tablecolumns{12}
\rotate
\tabletypesize{\scriptsize}
\tablecaption{{\it RXTE} Timing of Full \exo\ Eclipses\label{tbl:fitable}}
\tablewidth{0pc}
\tablehead{\colhead{ObsID} 
&\colhead{Cycle} 
&\colhead{Mid-Eclipse} 
&\colhead{Error} 
&\colhead{Ingress} 
&\colhead{Totality} 
&\colhead{Egress} 
&\colhead{Pre-Ingress}
&\colhead{Totality} 
&\colhead{Post-Egress}
&\colhead{Eclipse}
&\colhead{Number}\\
 &Number &Time & &Duration &Duration &Duration &Count Rate &Count Rate &Count Rate &Fit &of PCUs\\
 &  & (MJD;TDB) & (s) & (s) & (s) & (s) & (cnts s$^{-1}$) & (cnts s$^{-1}$) & (cnts s$^{-1}$) &${\chi_{\nu}}^2$ & }
\startdata
      10108-01-01-00 &25702 &50206.3747810 &0.13 &1.72(0.40) &508.26(0.44) &3.33(0.47) &112.6(1.3) &4.27(0.38) &123.6(1.4) &1.05 &5\\[0.008in]
      10108-01-02-00 &25703 &50206.5340983 &0.12 &1.29(0.29) &505.04(0.28) &2.82(0.34) &117.5(1.9) &2.90(0.35) &127.8(1.9) &1.07 &5\\[0.008in]
      10108-01-03-00 &25704 &50206.6934275 &0.086 &0.80(0.27) &504.35(0.42) &4.85(0.68) &107.3(1.8) &3.16(0.36) &120.8(1.8) &1.12 &5\\[0.008in]
      10108-01-04-00 &25705 &50206.8527675 &0.30 &3.46(0.65) &501.52(0.72) &11.01(0.83) &101.1(1.7) &3.71(0.36) &110.7(1.8) &1.16 &5\\[0.008in]
      10108-01-05-00 &25706 &50207.0121102 &0.28 &2.35(0.49) &503.14(0.48) &6.41(0.63) &99.4(1.7) &3.03(0.34) &105.1(1.8) &1.14 &5\\[0.008in]
      10068-03-02-00 &26278 &50298.1533303 &0.16 &3.44(0.59) &497.89(0.38) &5.15(0.53) &110.8(1.9) &-1.55(0.37) &138.7(2.0) &1.16 &5\\[0.008in]
      10108-01-06-00 &26353 &50310.1036665 &0.21 &1.86(0.61) &493.87(0.77) &10.9(1.4) &93.2(1.8) &7.12(0.44) &110.2(1.9) &1.21 &5\\[0.008in]
      10108-01-07-00 &26354 &50310.2630253 &0.31 &8.5(1.1) &493.38(0.72) &6.55(0.87) &121.5(1.9) &5.22(0.41) &136.7(2.0) &1.30 &5\\
\enddata
\tablenotetext{a}{On January 16, 2009, the RXTE Guest Observer Facility annnounced that
a one-second timing error had been discovered in the raw data files from observations
in the date range 1-12 January, 2006. 
The fitted mid-eclipse time for this ObsID has been corrected for this error.
Information regarding this error can be found at the web address http://heasarc.gsfc.nasa.gov/docs/xte/xhp\_new.html\#timingerror}
\end{deluxetable}
%
%
\begin{deluxetable}{ccl}
\tablecolumns{3}
\tablecaption{Unusable Observed \exo\ Eclipses\label{tbl:unfitable}}
\tablewidth{0pc}
\tablehead{\colhead{ObsID} 
&\colhead{Cycle} 
&\colhead{Comment}}
\startdata
10108-01-10-00      &26357  &Eclipse unusable; PCU4 switch-on with bad EDS mode.\\[0.008in]
20069-01-04-00      &27003  &Missing GoodXenon data file.\\[0.008in]
20069-05-05-00      &28335  &X-Ray burst during totality with decay after egress.\\[0.008in]
20082-01-03-05      &28688  &Ingress only in data.\\[0.008in]
20082-01-03-00      &28691  &Ingress only in data.\\[0.008in]
30067-04-03-00      &30999  &X-Ray burst during ingress; Eclipse profile corrupted.\\[0.008in]
40039-01-03-00      &31958  &Egress profile hidden by data gap.\\[0.008in]
40039-04-05-00      &32969  &X-ray burst before ingress distorts profile.\\
\enddata
\end{deluxetable}
%
%
\begin{deluxetable}{rcl}
\tablecolumns{3}
\tablewidth{0pc}
\tablecaption{Orbital Ephemerides of \exo\label{tbl:4epochephem}}
\tablehead{\colhead{Parameter} & & \colhead{Value}}
\startdata
\phm{X}$T_0$ (MJD/TDB)\phm{X}&$=$&$46111.0752010$\\[0.10in]
$\Porb_{,0}$ (day)&$=$&$0.15933772840$\\[0.10in]
$n_{b,0}$ (cycle)\phm{X}&$=$&$12526$\\[0.10in]
$T_{b,0}$ (MJD/TDB)\phm{X}&$=$&$48107$\\[0.10in]
$\Porb_{,1}$ (day)&$=$&$0.15933782601$\\[0.10in]
$\Porb_{,1}-\Porb_{,0}$ (ms)&$=$&$+ \, 8.43$\\[0.10in]
$n_{b,1}$ (cycle)\phm{X}&$=$&$34515$\\[0.10in]
$T_{b,1}$ (MJD/TDB)\phm{X}&$=$&$51610.7$\\[0.10in]
$\Porb_{,2}$ (day)&$=$&$0.15933775443$\\[0.10in]
$\Porb_{,2}-\Porb_{,1}$ (ms)&$=$&$- \, 6.18$\\[0.10in]
$n_{b,2}$ (cycle)\phm{X}&$=$&$45515.7$\\[0.10in]
$T_{b,2}$ (MJD/TDB)\phm{X}&$=$&$53363.4$\\[0.10in]
$\Porb_{,3}$ (day)&$=$&$0.15933783446 $\\[0.10in]
$\Porb_{,3}-\Porb_{,2}$ (ms)&$=$&$+ \, 6.91$\\[0.10in]
$\Porb_{,3}-\Porb_{,0}$ (Net,ms)&$=$&$+ \, 9.16$\\[0.10in]
\enddata
\end{deluxetable}

\clearpage

%
%
%
%
%
%
%
%
%
%

\end{document}